\documentclass[prd,aps,floats,tabularx,preprintnumbers,twocolumn]{revtex4}
\usepackage{graphicx, epsfig}

\def\be{\begin{equation}}
\def\ee{\end{equation}}
\def\ba{\begin{eqnarray}}
\def\ea{\end{eqnarray}}
\def\barr{\overline{r}}

\begin{document}

\baselineskip 12 pt

\title{Nontrivial Geometries:\\ Bounds on the Curvature of the Universe}

\author{Laura Mersini-Houghton$^\ast$, 
Yun Wang$^\dagger$,
Pia Mukherjee$^{a}$,  
and Ervin Kafexhiu$^{b}$}

\affiliation
{$^\ast$ UNC-Chapel Hill, CB{\#}3255, 
Phillips Hall,Chapel Hill, NC 27599, USA,\\ 
$^\dagger$ Dept. of Physics \& Astronomy, Univ. of Oklahoma,
440 W. Brooks St., Norman, OK 73019\\
$^{a}$ Department of Physics \& Astronomy, Univ. of Sussex, 
                 Falmer, Brighton, BN1 9QH, UK\\ 
 $^{b}$Department of Physics and Astronomy,\\ Physics Department, University of Tirana, Tirana, Albania}

\begin{abstract}
{\small}
%%%%%%%
Probing the geometry of the universe is one of the most important endevours in cosmology. 
Current observational data from the Cosmic Microwave Background 
anisotropy (CMB), galaxy surveys and type Ia supernovae (SNe Ia) 
strongly constrain the curvature of the universe to be close to zero 
for a universe dominated by a cosmological constant
or dark energy with a constant equation of state. 
Here we investigate the role of cosmic priors on deriving
these tight bounds on geometry, by considering a landscape motivated
scenario with an oscillating curvature term. We perform a
likelihood analysis of current data under such a model of
non-trivial geometry and find that the uncertainties on curvature, 
and correspondingly on parameters of the matter and dark energy sectors,
 are larger. Future dark energy experiments together with CMB data from 
experiments like Planck could dramatically improve our ability to 
constrain cosmic curvature under such models enabling us to probe 
possible imprints of quantum gravity.

\end{abstract}

\maketitle

%%%%%%%%%%%%%%%%%%%%%%%%%%5555
\section{\bf Introduction}

Exquisite measurements of the Cosmic Microwave Background (CMB) 
spectra, Large Scale Structure (LSS) and of the expansion history of the universe, by various experiments including WMAP, SDSS, 2dF survey, and 
SN1a surveys \cite{wmap, sdss, 2df, riess, perl} 
have pinned down the spatial curvature of the universe to $\kappa 
\le 0.01$ \cite{Spergel06, Tegmark06} in a universe with a 
cosmological constant .

Energy content and geometry both contribute to the Hubble expansion rate $H$. The 
three are so closely intertwined that no independent measurement of $H$ versus 
$\Omega_m$, $\Omega_{\Lambda}$, or $\kappa$ can yet be performed. In interpreting the data for any of these components, we have to be
aware of the prior assumptions in our models regarding the other components. That is to say that our interpretation of data is not model independent.

What conclusion can be made about the geometry of the universe, on the basis 
of current data? The answer to this question depends on the cosmological model and on the dependence of the constraints on the spatial geometry of 
the universe on prior assumptions regarding other relevant energy components. Here we explore this issue by considering a model of nontrivial geometry,  
where the curvature does not take a constant value, but rather is a 
function of time. In our model the curvature is given by an oscillating function with a Hubble time period. Such models can be motivated by the dissipative dynamics of the wavefunction of the universe on its classical path
on the background of the landscape of string theory.

%The motivation for the phenomenological scenario with a time-dependent 
%curvature is twofold: first, we would like to inquire on the 
%robustness of the evidence for a flat universe and a concordance 
%cosmology from current data in the light of degeneracies with 
%other cosmic priors; second, theoretically, models with nontrivial 
%oscillating curvature can be motivated by the dissipative dynamics 
%of the wavefunction of the universe on its classical path
%on the background of the landscape of string theory.

\section{\bf Motivating the Class of Oscillating Curvature Models}

The discovery of the acceleration of the universe has become one of the central 
themes of current investigation in physics. Unfortunately due to the degeneracy 
among cosmic parameters, determining its nature and equation of state from 
astrophysical data depends crucially on our assumptions for 
the matter content and curvature of the universe. Recent analysis of SDSS\cite{sdss} 
and WMAP \cite{Spergel06} has been reported to indicate that the geometry of our universe 
is extremely close to flat to within $1\%$ in an LCDM universe \cite{Tegmark06}.   

Here we would like to investigate the robustness of these conclusions about the 
curvature of the universe by presenting a highly nontrivial model where the 
{\it curvature term is not a constant but a function of time, oscillating  
every Hubble time}. This model is inspired and motivated from the proposal 
for a dynamic selection of the initial conditions for our universe from the 
landscape phase space \cite{laurarich1,laurarich2} as summarized below.

In \cite{laurarich1,laurarich2} we included the backreaction of superhorizon 
massive perturbations on the initial wavepacket for our universe. Solving a 
Master Equation we studied role of the backreaction term on the decoherence 
of our initial patch from the other WKB branches 
on the landscape. The time evolved nonlocal entanglement of our 
patch with others outside the horizon at late times \cite{laurarrow,kiefer} 
were then investigated in \cite{lauratomo} with the
conclusion that some of those traces imprinted on CMB and LSS are within 
observational reach. Within this formalism, we now allow for an initial 
curvature term in the Friedman equation, 
$\kappa=1$ and consider the effect that the backreaction term has 
on closed geometries. Backreaction shifts the energy of the wavepacket 
therefore its classical trajectory on the landscape. The shifting of 
the classical path for our wavefunction can be seen  by integrating out the 
Master Equation presented for this case as was described in 
\cite{laurarich1,lauratomo,laurarrow}. 

\begin{equation}
H(a) + H_m =\sum_{n} H_n
\label{master}
\end{equation}

where $-H(a)=a ( \dot{a}^2 +\kappa )$ is the hamiltonian corresponding 
to gravitational degrees of freedom (obtained from the Einstein-Hilbert action), with $a$ the scale factor, and $\kappa$ is the initial 
curvature for that classical trajectory. $H_m$ is the matter hamiltonian corresponding for example to 
the inflaton energy and, $H_n$ is the backreaction energy corresponding 
to the superhorizon matter perturbations labelled by the wavenumber $n$.

To get a rough idea of the shifting of the trajectory, let us assume 
that $\epsilon = a^3 [\dot{\phi}^2 + m^2 \phi^2]$ is a 
constant of motion and thus integrate out Eqn.\ref{master}. When the 
backreaction term is not included, Eqn.\ref{master} in the case of closed universes 
$\kappa =1$, 
gives a turning point when $\dot{a}=0$ at $a=a_n$ where $a_n \kappa=\epsilon$. 
The first term in the backreaction energy $H_n \simeq n^2 / a^2$ has the same 
dependence on the scale factor $a$ as the curvature. Including the backreaction 
term when integrating out Eqn.\ref{master} results in a lower energy since 
$H(a) \rightarrow H(a)-\sum H_n$, thereby shifting the classical trajectory 
of the wavepacket of our universe. This result of the shifting of the classical 
trajectory of our universe's wavepacket by interaction with a field, (in this 
case the entanglement of gravitational and matter degrees of freedom through 
the term $\sum H_n$), is well known in particle physics where the energy of a 
quantum particle gets shifted by interaction with a classical field which 
results in a shifts of the particle's trajectory and momenta.
The details of the calculations for the strength of this interaction in the 
case of quantum cosmology can be found for example in \cite{kiefer} or when 
applied to the landscape of string theory in \cite{lauratomo}. The result 
is that everytime a closed universe goes through its turning point given by 
Eqn.\ref{master} by putting $\dot a =0$, that is every Hubble time, then it 
will emerge through a shifted trajectory, due to the correction of $H(a)$ 
by the backreaction term $\sum H_n$ described here. The modification in 
energy corresponding to this shift can be absorbed  into the curvature term 
since the time dependence of $H_n$ is similar to that of the $\kappa$ term. 
Therefore, to local observers bound to our visible universe, the effect of 
the shifting of the classical trajectory for our universe in the phase space, 
appears as an induced oscillating curvature with a period of Hubble time. 
Of course no observers would survive the emergence through the turning point 
in the cycles of the trajectory. However the reduced oscillation of the 
curvature in the previous cycle may leave its imprints on astrophysical 
observables of the current cycle, which we aim to study here. 
Based on the integration of Eqn.\ref{master}, we expect 
the curvature term motivated by this scenario, to be a function of the 
total energy content of the universe $H_m$ and oscillate, in each cycle between the  
two turning points in the trajectory, with a Hubble period.
%We present such an oscillating function for the curvature below, Eq.\ref{oscicurv}. 
%The Hubblee dependence in the oscillating function has been replaced by an equivalent 
%expression given in terms of the redshift $z$. The latter replacement is simply meant 
%to simplify the technical details of the likelihood analysis but otherwise carries 
%the same functional dependence.

While there are different ways of phenomenologically implementing 
the oscillating curvature model described above, there
are constraints based on the considerations of observables
in the model.
In order to make predictions that can be compared with data,
we need to compute the coordinate distance $r(z)$ and the 
age of the universe $t(z)$ as functions of redshift.

These functions, $r(z)$ and $t(z)$, are found by considering 
the radial, null geodesic of the Robertson-Walker metric. 
But now the curvature constant $\kappa$ is replaced by a function
$-x(a)$ that oscillates with cosmic time:
\begin{equation}
ds^2=dt^2 - a^2(t)\, \frac{dr^2}{1+ x(a) \,{\barr}^2 } = 0,
\end{equation}
where $a(t)$ is the cosmic scale factor, and
the scaled coordinate distance $\barr \equiv H_0 r/c$.
Hence we have
\begin{equation}
\left(  \barr^\prime \right)^2 = \frac{1+ x(a)\,{\barr}^2 }{a^4 E^2(a)},
\label{eq:drda}
\ee
where $\prime$ denotes differentiation with respect to $a$, and
\be
E^2(a) \equiv \left(\frac{H(z)}{H_0}\right)^2 =
\frac{\Omega_r}{a^4} + \frac{\Omega_m}{a^3} +\Omega_\Lambda+ 
\frac{x(a)}{a^2}.
\ee
For closed universes, $r(z)$ is not a monotonic function of
$a$. Differentiating Eq.(\ref{eq:drda}) with respect 
to $a$ gives
\be
\barr^{\prime\prime}=\frac{x^\prime \barr^2 + \barr^\prime 
\left[2\barr x-\barr^\prime \left( \Omega_m+4a^3 \Omega_\Lambda+ 
 x^\prime  a^2 +2 a x\right)\right]}{2 a^4 E^2(a) \,\barr^\prime }.
\label{eq:r(z)eq}
\ee
In a closed universe, the coordinate distance $r$ reaches
its maximum value at $\barr^\prime=0$. 
Note that $\barr^{\prime\prime}$ is only finite at 
$\barr^\prime=0$ if and only if $x^\prime=0$ at $\barr^\prime=0$.
This can only be satisfied if $x=$constant (the usual constant
curvature case), or $x=x(r)$.

Based on the dissipative dynamics of the shifted cycles of the universe 
described above, we consider the following heuristic model that captures 
the desired features and satisfies the above constraint 
\be
x(a)  \equiv  \frac{\Omega_k(z)}{(1+z)^2} =\Omega_k-A\sin(B\barr),
\label{oscicurv}
\ee
where $\Omega_k\equiv \Omega_k(0)=1-\Omega_r-\Omega_m-\Omega_\Lambda$,
$\Omega_r$, $\Omega_m$, $\Omega_\Lambda$ denote 
the present day density fractions of radiation, matter, and vacuum energy,
and $r$ denotes the coordinate distance from the observer at $z=0$
to redshift $z$. $A$ and $B$ are dimensionless constants.  
Note that the conventional model with constant curvature is recovered for $A=0$,
$x(a)=\Omega_k(z)/(1+z)^2=\Omega_k=-\kappa/H_0^2$, where $\kappa$ is the curvature
constant.

Except for the special case of constant curvature ($A=0$),
$r(z)$ is found by numerically solving the second order 
differential equation in Eq.(\ref{eq:r(z)eq}),
with the initial condition that at $a=1$, $\barr=0$, $\barr^\prime=-1$.

Fig.1 shows models with $B=0.5$, 1, and 2 respectively.
Fig.2 shows $r(z)$ for the models in Fig.1,
with the same line types. 
\begin{figure} 
\psfig{file=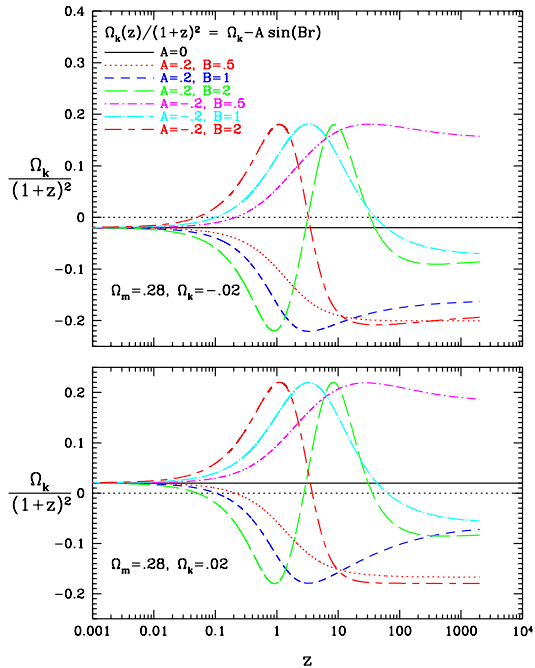,width=3.6in}
\caption{Oscillating curvature models with $B=0.5$, 1, and 2.}
\end{figure}
\begin{figure} 
\psfig{file=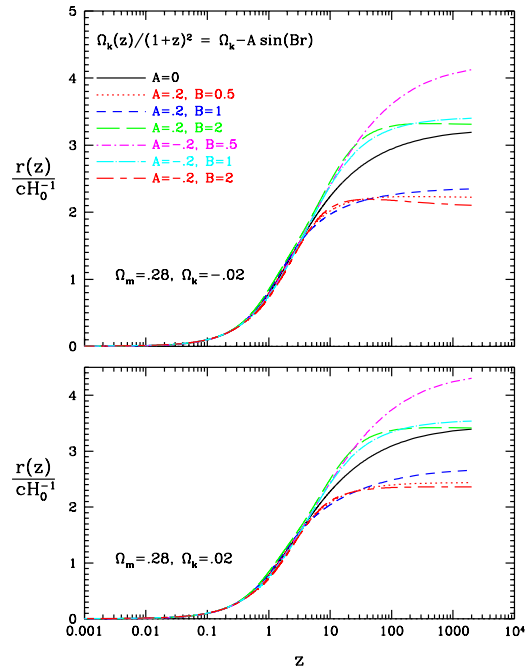,width=3.6in}
\caption{The coordinate distance $r(z)$ for oscillating curvature models 
in Fig.1, with the same line types.}
\end{figure}

The age of the universe is given by
\be
t(a)= H_0^{-1} \int_0^a \frac{da}{a\, E(a)},
\ee
once $\Omega_k(z)$, which now depends on $z$ through $r(z)$, has been found numerically.

\section{Observational constraints on oscillating curvature and the energy content}

We use current observational data to constrain
the oscillating curvature model given by Eq.(\ref{oscicurv}).
Following the approach of \cite{WangPia07}, 
we assume the HST prior of $H_0=72\pm 8\,$(km/s)Mpc$^{-1}$ \cite{HST},
use 182 SNe Ia (from the HST/GOODS program \cite{Riess07}, 
the first year Supernova Legacy Survey \cite{Astier05}, 
and nearby SN Ia surveys) \cite{Riess07}, 
CMB data \cite{Spergel06}, and the SDSS measurement of the baryon 
acoustic oscillation scale \cite{Eisen05}.
We use the CMB data in the form of the 
CMB shift parameters $R\equiv \sqrt{\Omega_m H_0^2} \,r(z_{CMB})$
and $l_a\equiv \pi r(z_{CMB})/r_s(z_{CMB})$ derived from WMAP three year
data by \cite{WangPia07}.

We run a Markov Chain Monte Carlo (MCMC) \cite{Lewis02} to obtain ${\cal O}$($10^6$) samples for each set of 
results presented in this paper. The chains are subsequently 
appropriately thinned.

Due to the degeneracies between $A$, $B$, and $\Omega_k$,
($A$, $B$, $\Omega_k$) are not well constrained when they
are all allowed to vary. To illustrate the effect of
oscillating curvature, let us study
the class of models given by Eq.(\ref{oscicurv}) for fixed representative values of $B$,
while allowing $A$ and $\Omega_k$ to vary, along with $\Omega_m$,
$\Omega_b h^2$, and $h$ (see \cite{WangPia07}).
The parameters estimated from data are ($\Omega_m$,
$\Omega_b h^2$, $h$, $\Omega_k$, $A$). It should be noted that $A$ affects the 
overall amplitude of the curvature term, while $B$ plays the role of its oscillating 
frequency. The case $B=1$ would correspond to $\Omega_k$ oscillating every Hubble time.

Figs.{\ref{fig:Bd5_omokA}}-{\ref{fig:B5_omokA}} show the joint
confidence contours in the plane ($\Omega_m$, $\Omega_k$)
and ($\Omega_k$, $A$) for $B=0.5$, 1, 2, and 5 respectively.
The inner and outer contours correspond to 68\% and 95\%
confidence levels respectively.

\begin{figure} 
\psfig{file=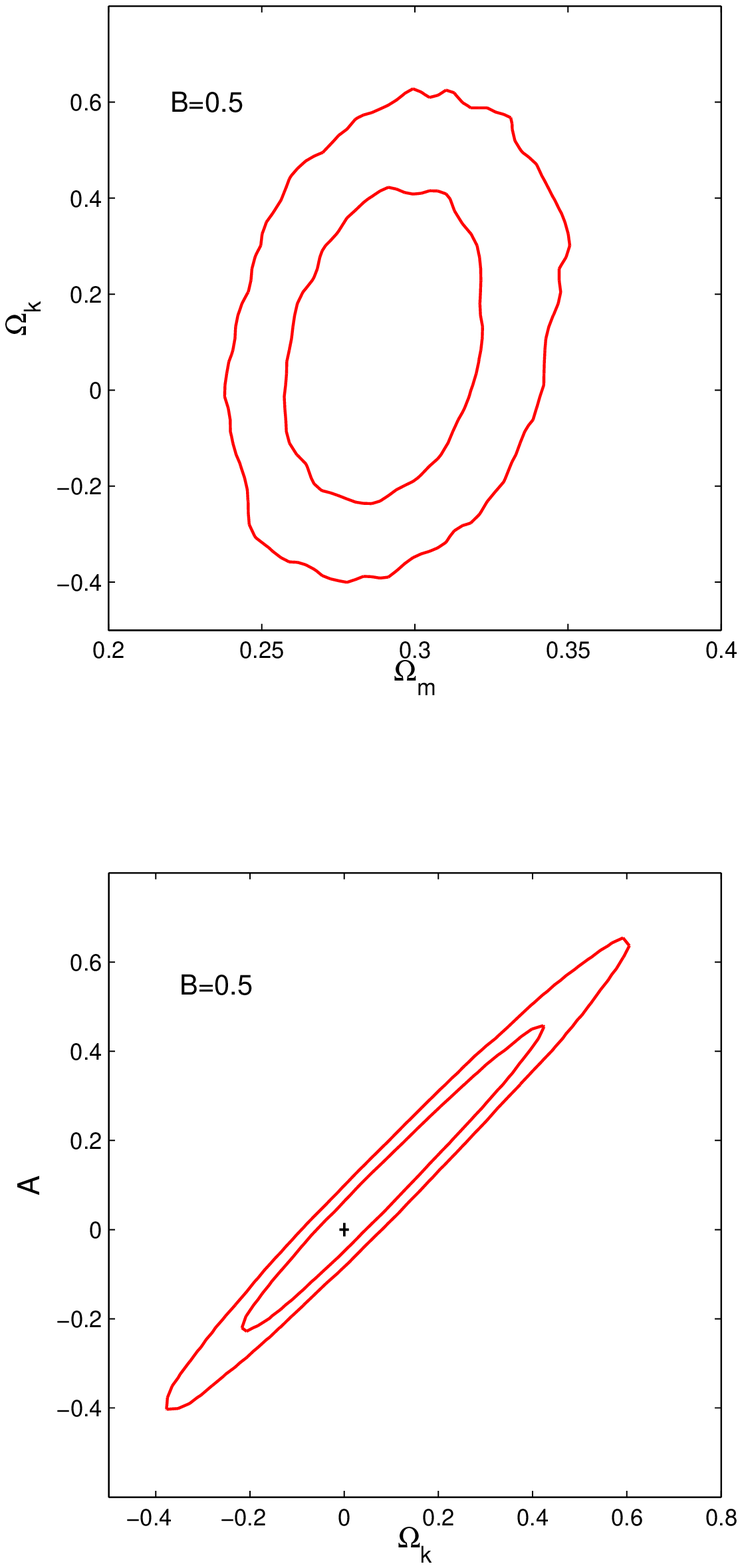,width=2.8in}
\caption{Joint confidence contour plots for ($\Omega_m$, $\Omega_k$)
and ($\Omega_k$, $A$) for $B=0.5$.
The inner and outer contours correspond to 68\% and 95\%
confidence levels respectively.}
\label{fig:Bd5_omokA}
\end{figure}

\begin{figure} 
\psfig{file=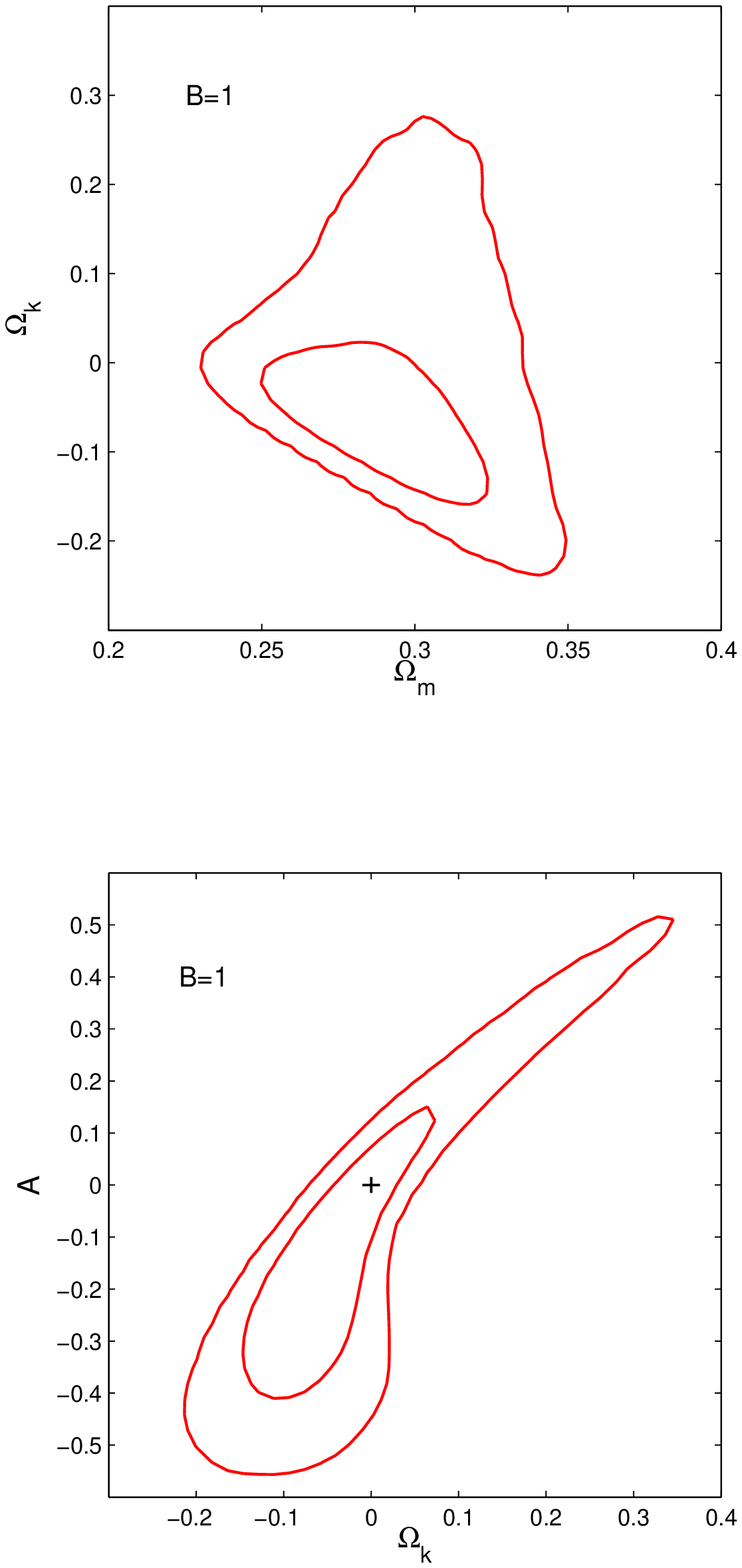,width=2.8in}
\caption{Joint confidence contour plots for ($\Omega_m$, $\Omega_k$)
and ($\Omega_k$, $A$) for $B=1$.
The inner and outer contours correspond to 68\% and 95\%
confidence levels respectively.}
\label{fig:B1_omokA}
\end{figure}

\begin{figure} 
\psfig{file=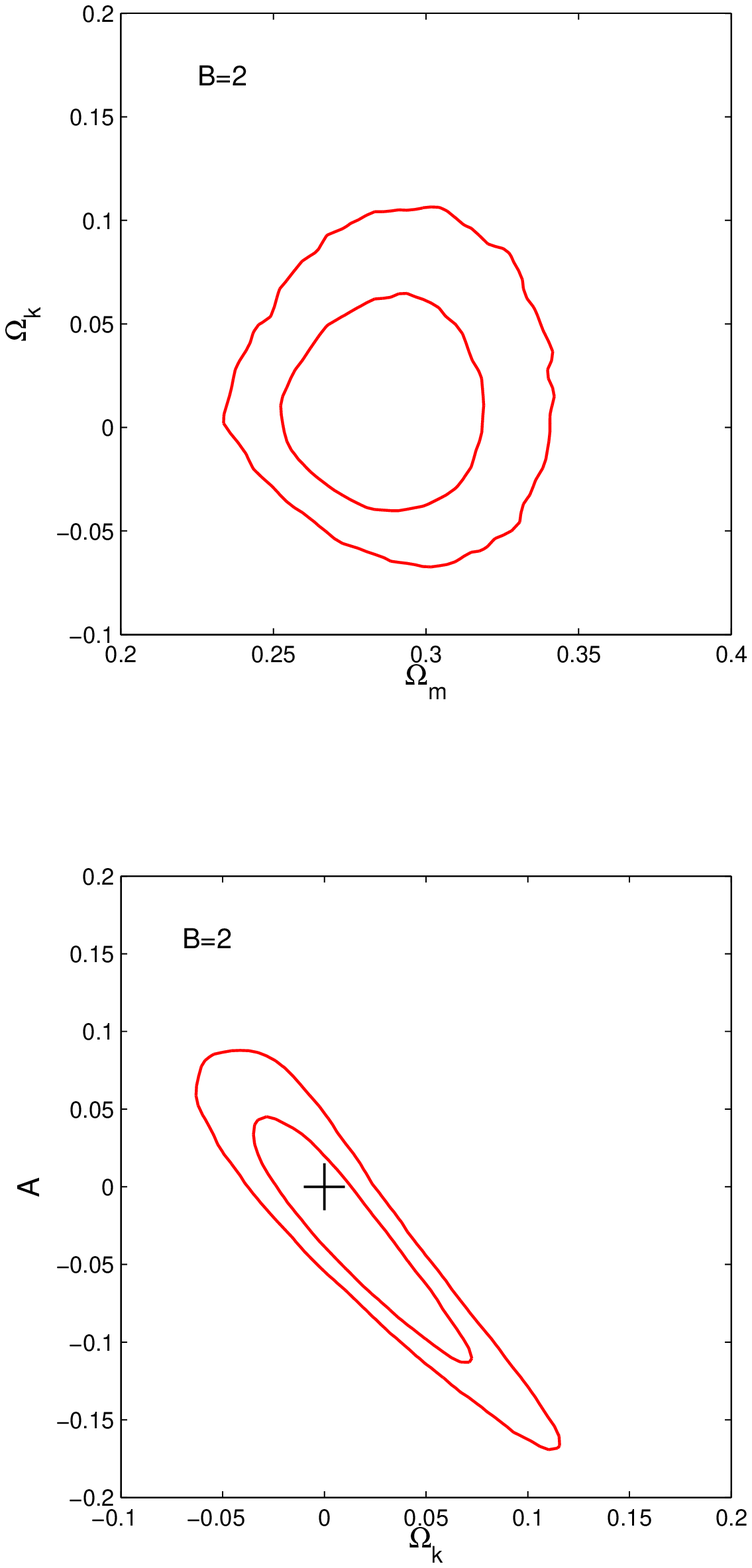,width=2.8in}
\caption{Joint confidence contour plots for ($\Omega_m$, $\Omega_k$)
and ($\Omega_k$, $A$) for $B=2$.
The inner and outer contours correspond to 68\% and 95\%
confidence levels respectively.}
\label{fig:B2_omokA}
\end{figure}

\begin{figure} 
\psfig{file=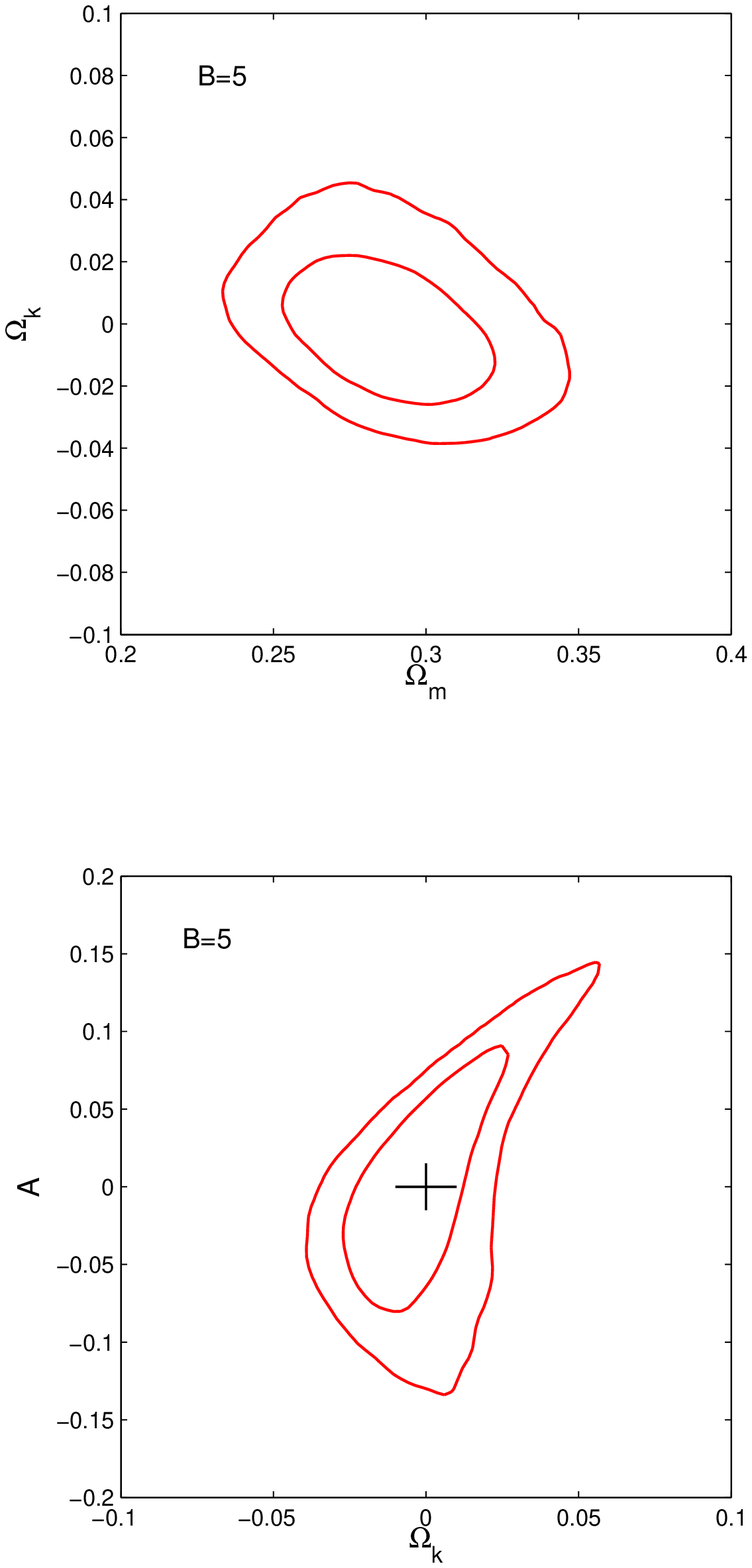,width=2.8in}
\caption{Joint confidence contour plots for ($\Omega_m$, $\Omega_k$)
and ($\Omega_k$, $A$) for $B=5$.
The inner and outer contours correspond to 68\% and 95\%
confidence levels respectively.}
\label{fig:B5_omokA}
\end{figure}

As can be seen from the plots given in 
Figs.{\ref{fig:Bd5_omokA}}-{\ref{fig:B5_omokA}}, 
current data allow models in which the curvature of the
universe oscillates with cosmic time. 
The allowed range of the current curvature density ratio $\Omega_k$
is significantly increased compared to the case of
constant curvature.

The bounds derived from the WMAP three year data and galaxy survey data from 
the SDSS \cite{sdss} give for the case of constant curvature,  
$\Omega_k = -0.005 \pm 0.006$, (2dF data \cite{2df} also give similar results) 
\cite{Spergel06}. 
Comparing these bounds to the case of oscillating curvature models, we find
that the constraints on the geometry of the universe change significantly,
now we have $\Omega_k = 0.097\pm  0.210$ for $B=0.5$,
$\Omega_k = -0.037\pm 0.092$ for $B=1$,
$\Omega_k = 0.014\pm  0.035$ for $B=2$,
and $\Omega_k = 0.000\pm 0.017$ for $B=5$.
The constraints on the $\Omega_k$ and $A$ become more
stringent as $B$ increases. This is as expected,
since $B$ is the curvature oscillation frequency.
For large $B$, the cumulative effect of the oscillating
curvature decreases. It is very interesting that when the period of the 
curvature oscillation $B^{-1}$ becomes larger than a Hubble time, 
the range of the allowed values for $\Omega_k$ and $\Omega_m$,
at $95\%$ confidence level in agreement with data, shows a drastic 
increase. An oscillation in the curvature with a period larger than the 
age of the universe, a case which locally would appear as nearly a 
constant while being globally notrivial, the time dependence of which 
would otherwise not be captured by data, does in fact contain a significant 
deviation from the priors of a simple $LCDM$ model with constant or zero curvature. 
This is one of our important results: a highly nontrivial geometry on 
scales larger than the horizon can lead to a very different interpretation of data.

Let us now investigate the implications of the oscillating curvature 
for the dark energy equation of state $w$. 
This is done by conducting a likelihood analysis using MCMC of the
oscillating curvature model, Eq.(\ref{oscicurv}),
assuming a constant dark energy equation of state $w_X(z)=w$.
The parameters estimated from data are ($\Omega_m$,
$\Omega_b h^2$, $h$, $\Omega_k$, $A$, $w$). 
Fig.{\ref{fig:B5_w_omokA}} shows the
joint confidence contours of ($\Omega_m$, $\Omega_k$)
and ($\Omega_k$, $A$), and Fig.{\ref{fig:B5_w_okw}} 
shows the joint confidence contour of
($\Omega_k$, $w$), for $B=5$.
As expected, adding $w$ as an additional parameter
to be estimated from data notably increase the uncertainties
on estimated parameters, especially ($\Omega_k$, $A$, $w$).
For example, for $B=5$, $\Omega_k =0.074\pm  0.068$ when $w$
is included as an estimated parameter, compared 
with $\Omega_k = 0.000\pm 0.017$ for setting $w=-1$
(a cosmological constant).
For larger values of the period $B^{-1}$, the uncertainties on
($\Omega_k$, $A$, $w$) increase significantly. Notice that the bounds on $w$, 
as shown in Fig.8 for the case $B=5$ can be as large as 
$-1.04 \le w \le -0.71$ at $95\%$ confidence level. 
These bounds on $w$ should be contrasted to the constraints derived 
in \cite{melchiorri} where a prior of $\Omega_k =0$ was assumed. 
Clearly, current precision cosmology data is not sufficient in 
pinning down the equation of state for dark energy when the 
geometry of the universe is nontrivial.
 
\begin{figure} 
\psfig{file=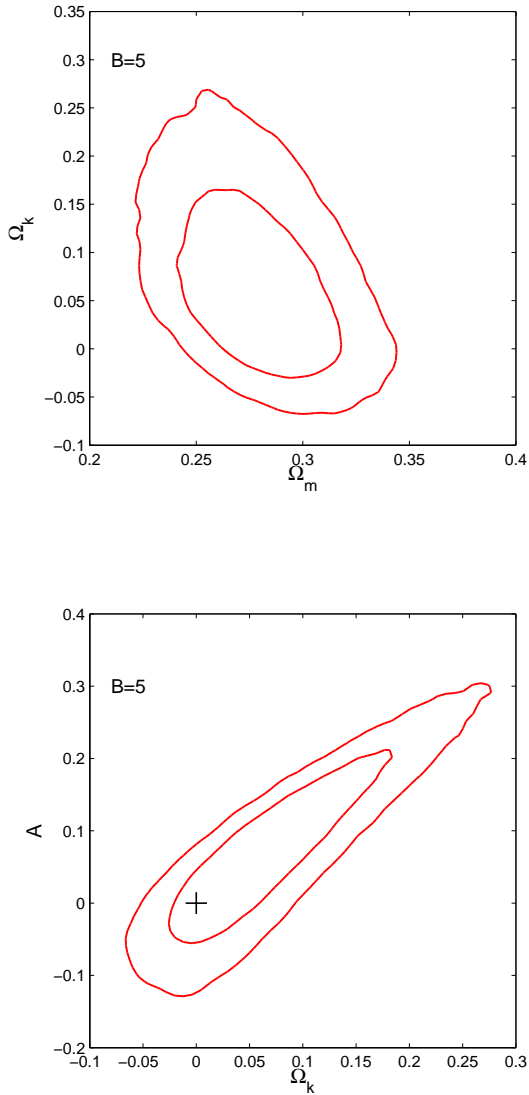,width=2.8in}
\caption{Joint confidence contour plots for ($\Omega_m$, $\Omega_k$)
and ($\Omega_k$, $A$) for $B=5$.
The inner and outer contours correspond to 68\% and 95\%
confidence levels respectively.}
\label{fig:B5_w_omokA}
\end{figure}

\begin{figure} 
\psfig{file=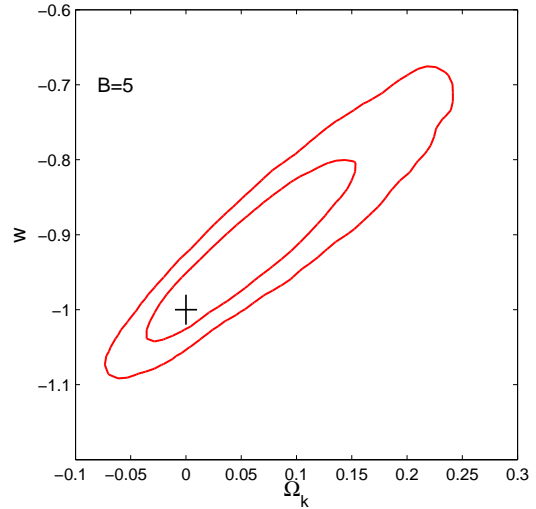,width=2.8in}
\caption{Joint confidence contour plots for ($\Omega_k$, $w$)
for $B=5$.
The inner and outer contours correspond to 68\% and 95\%
confidence levels respectively.}
\label{fig:B5_w_okw}
\end{figure}

\section{Discussion}
We have studied constraints on the parameters of a
landscape motivated cosmological model in which the
curvature of the universe oscillates with cosmic time
(see Eq.(\ref{oscicurv})). Such a model is motivated from the proposal 
for a dynamic selection of the initial conditions for our universe from 
the landscape phase space. Thus an analysis of the kind performed here 
could lead to the implicit detection of quantum gravity effects.

We have used CMB data in the form 
of the shift parameters $R$ and $l_a$ extracted from WMAP three year data 
by \cite{WangPia07}, together with the SDSS measurement of baryon acoustic 
oscillation scale \cite{Eisen05}, and SN Ia data from HST and ground-based 
observations \cite{Riess07,Astier05}. From the bounds derived on the 
parameters of this model we find that currently a
simple flat model, which is a special case of the above model, remains
a good bet; such a conclusion will be supported further by model selection 
arguments \cite{Pia06}.
Allowing for nontrivial geometry leads to greater 
uncertainties in our knowledge of the present 
day curvature and matter density ratios $\Omega_k$ and $\Omega_m$, as can 
be seen in Figs.{\ref{fig:Bd5_omokA}}-{\ref{fig:B5_w_okw}}).
An oscillating curvature term also significantly
changes the bounds on the dark energy equation of state
$w$, as seen in Fig.{\ref{fig:B5_w_okw}}.

It would be interesting to look for the imprints of such a model as data get
 better. Future dark energy experiments from both ground and space
\cite{Wang00a,detf,ground,jedi}, together with CMB data from 
Planck \cite{planck},
should dramatically improve our ability to constrain cosmic curvature,
and probe possible imprints of quantum gravity.

{\it Acknowledgements} L.M.H is supported in part by DOE grant 
DE-FG02-06ER41418 and NSF grant PHY-0553312. Y.W is supported 
in part by NSF CAREER grant AST-0094335 (YW).
PM is supported by PPARC, UK.

\end{document}